\documentstyle[12pt,epsfig]{article} 
\begin{document}
\title{Time Evolution of a Quantum Particle and a Generalized Uncertainty Principle}
\author{ A. Camacho 
\thanks{email: acamacho@janaina.uam.mx} \\
Physics Department, \\
Universidad Aut\'onoma Metropolitana-Iztapalapa. \\
P. O. Box 55-534, 09340, M\'exico, D.F., M\'exico.}
\date{}
\maketitle

\begin{abstract}
Some of the possible consequences of a generalized uncertainty principle (which emerges in the context of string theory and quantum gravity models as a consequence of fluctuations of the background metric) are analyzed consi\-dering the case of a quantum particle immersed in a homogeneous gravitational field. It will be shown that the expectation value of the momentum operator depends in a novel way on the mass of the involved particle. This kind of physical characteristics could be, in principle, detected. In other words, one way of confronting against the experiment some of the models around quantum gravity is given by the detection of the dependence upon the mass parameter of the expectation value of the momentum operator.

\end{abstract}
\bigskip

\section{Introduction}
\bigskip

One of the most difficult tasks of modern physics comprises the reconciliation of quantum theory (QT) with general relativity (GR).  Though this is still an open pro\-blem, there are already some models that appear as feasible candidates. Among them we may mention string theory and quantum gravity, models that entail a modification of the well known uncertainty principle [1, 2]. The new form of this principle is closely related to the fact that the conventional notion of distance breaks down somewhere around the Planck distance, $L_p = \sqrt{{\hbar G\over c^3}}$. This generalized uncertainty principle (GUP) implies also the correction of the commutation relation in the corresponding Heisenberg algebra [3].

The aforementioned change has also been derived not only in connection with quantum geometry [4], but also in association with the way in which quantum measurements alter the local spacetime metric [5]. The emergence of a GUP in several models has suggested that it could be a very general characteristic shared by several theories [6]. This last statement means that the analysis of the consequences of this kind of GUPs could render important conclusions about the physical implications of the attempts to quantize the gravitational field.

In the present work the relation between one of the possible GUPs and the time evolution of a quantum particle immersed in a homogeneous gravitational field will be analyzed. The motion equations for the momentum and position ope\-rators will be solved, and it will be shown that the momentum operator does depend in a novel way  upon the mass of the involved particle. This kind of physical characteristics could be, in principle, detected. In other words, one way of confronting against the experiment some of the models around quantum gravity is given by the detection of the dependence upon the mass parameter of the expectation value of the momentum operator.

\bigskip

\section{GUP and the motion equations in a homogeneous gravitational field}
\bigskip

Let us now assume that the Heisenberg algebra takes the form [3, 4, 5]

\begin{equation}
[\hat{x}, \hat{p}] = i\hbar\Bigl[\Pi + \epsilon({L_p\over\hbar})^2\hat{p}^2\Bigr]. 
\end{equation}

Concerning this last expression it must be mentioned that $\epsilon$ denotes a constant, additionally we have considered a one--dimensional system, of course, $\Pi$ denotes the unit operator. The reason for this last restriction stems from that fact that up to now there is no higher--dimensional gene\-ralization of (1) [3].

Consider now a very simple situation, namely a quantum particle of mass $m$ obeys the following Hamiltonian

\begin{equation}
H = {p^2\over 2m} - mgx. 
\end{equation}

 Clearly the involved gravitational potential is related to a homogeneous field. 
This is a very simple situation and it may be argued that it provides a very rough a\-pproximation to any realistic physical scenario. Nonetheless, we may justify the use of such a simple gravitational field as follows: the emergence of GUPs is a dynamical phenomenon, namely, it is related to the existence of fluctuations of the background metric [4]. Hence, we may assume that in our context the background metric has been approximated by a weak field and slow motion limit, such that we may consider a Newtonian situation, and introduce the effects of the fluctuations of the metric, upon the dynamics of a nonrelativistic particle, taking into account a GUP in the motion equation for the corresponding operators. 

Clearly, the assumption of a Newtonian limit is a very stringent restriction. No\-netheless, we may introduce it as the roughest approximation for the background geometry (but the one allows us to deduce an analytical solution to the motion equations), and consider the presence of GUP as quantum gravity corrections, associated with fluctuations of the spacetime geometry, that in our case has an average geometry provided by a homogeneous Newtonian gravitational field. In other words, the fluctuations in the geometry are introduced only in the modifications that they provoke in the Heisenberg algebra.
Of course, a more realistic scenario has to consider a better approximation for the background geometry, but the present approach will give us a taste of the conceptual problems that might emerge if we consider, simultaneously, a gravitational field and a GUP, as fundamental elements in the time evolution of a quantum particle.

Hence, in the Heisenberg picture [7] the motion equations for the momentum and position operators read

\begin{equation}
{d\hat{p}\over dt} = mg\Bigl[\Pi + \epsilon({L_p\over\hbar})^2\hat{p}^2\Bigr], 
\end{equation}

\begin{equation}
{d\hat{x}\over dt} = {1\over m}\Bigl[\hat{p} + \epsilon({L_p\over\hbar})^2\hat{p}^3\Bigr]. 
\end{equation}
\bigskip

The solutions to these equations are, to first order in $\epsilon({L_p\over \hbar}\Bigr)^2$,

\begin{equation}
\hat{p}(t) = mgt\Pi + \hat{p}(0) + mg\epsilon\Bigl({L_p\over \hbar}\Bigr)^2\Bigl[{(mg)^2\over 3}t^3\Pi + mgt^2\hat{p}(0) + 
t\hat{p}(0)^2\Bigr],
\end{equation}

\begin{equation}
\hat{x}(t) = {gt^2\over 2}\Pi + {t\over m}\hat{p}(0) + \hat{x}(0)+ g\epsilon\Bigl({L_p\over \hbar}\Bigr)^2\Bigl[{(mg)^2\over 3}t^4\Pi + {4\over 3}mgt^3\hat{p}(0) + 2t^2\hat{p}(0)^2\Bigr].
\end{equation}

Here $\hat{p}(0)$ and $\hat{x}(0)$ are the initial momentum and position operators, respectively, which satisfy the commutation relation

\begin{equation}
[\hat{x}(0), \hat{p}(0)] = i\hbar\Bigl[\Pi + \epsilon({L_p\over\hbar})^2\hat{p}(0)^2\Bigr]. 
\end{equation}

Resorting to the aforementioned solutions, and to the last commutation relation, it is readily seen that

\begin{equation}
[\hat{x}(t), \hat{p}(t)] = i\hbar\Bigl[\Pi + \epsilon({L_p\over\hbar})^2\hat{p}(t)^2\Bigr]. 
\end{equation}

From the last expression we have that our GUP is always fulfilled. At this point it is noteworthy to comment that one of the consequences of a GUP like (1) is the fact that no physical state is a position eigenstate [3]. Notice that in our approach we have not considered position eigenstates, instead, we deal with position and momentum operators, i.e., we use Heisenberg picture. 

Consider now a quantum system whose vector state $\vert \alpha>$ is an eigenvector of the momentum operator, for our GUP this condition is, mathematically, consistent [3]. 

\begin{equation}
{<\alpha\vert \hat{p}(t)\vert\alpha>\over m} = {P(0)\over m} + gt\Bigl\{1 + \epsilon\Bigl({L_p\over \hbar}\Bigr)^2\Bigl[{(mgt)^2\over 3}\Pi + mgtP(0) + P(0)^2\Bigr]\Bigr\}.
\end{equation}

Here $P(0) = <\alpha\vert \hat{p}(0)\vert\alpha>$. At this point it is noteworthy to comment that we have imposed no restriction upon $\vert\alpha>$, therefore we may consider that it is related to the center of mass of a macroscopic body. Clearly (9) contains a novel mass dependence of the expection value. Indeed, $<\alpha\vert p(t)\vert\alpha>/m$ is not mass independent. In the limit $\epsilon\rightarrow 0$, the last two expressions, show that the Newtonian situation is recovered. 

Here we must add that there are further generalizations of (1) [3] which include a term with the form $\beta(x/L_p)^2$, and we could wonder about the possible effects of this kind of terms on the dynamics of our particle. Taking a look at (3) we may understand (on classical grounds) the last term on the right--hand side as an additional force, and the introduction of $\beta\not =0$ will not cancel, in the general case, the effects of $\epsilon(L_p/\hbar)^2p^2$. In other words, a deeper modification of the Heisenberg algebra would have similar consequences upon the dynamics of our quantum system.

\section{Conclusions}
\bigskip

We have considered a quantum particle moving in a region where a homogeneous gravitational field  is present, this field is introduced as an average geometry, and the fluctuations of the geometry have been introduced only in context of the modifications that they provoke in the commutation relations. Assuming a GUP, it was proved that $<\alpha\vert p(t)\vert\alpha>/m$ is, in the general case, not mass independent, the consequences of this last fact could be far--reaching. 
Clearly, expression (9) shows that WEP is not fulfilled. Here, following [8], WEP reads: {\it if an uncharged test body is placed at an initial event in spacetime and given an initial velocity then its subsequent trajectory will be independent of its internal structure and composition}. In other words, the presence of a GUP may be understood as an additional force in the dynamics of our particle. This last remark may be rephrased stating that one way of confronting against the experiment some of the models around quantum gravity is given by the detection of the dependence upon the mass parameter of the expectation value of the momentum operator.

Of course,  we must also address the issue concerning the feasibility of the detection of this type of effects. From (9) it is readily seen that the detection depends, critically, upon the magnitude of the term $\epsilon\Bigl({L_p\over \hbar}\Bigr)^2$. 
Here we face, from the very outset severe problems: 

Firstly, the magnitude of $\epsilon$, which here is a free parameter, can not be predicted from model--independent arguments. In other words, in order to have a value for this paramater we must consider a particular model [3], either string theory [9] (where the aforementioned parameter is related to the Regge slope), quantum geometry [4] (here the parameter is a function of an upper bound associated to the proper acceleration experienced by massive particles along their worldlines), etc., etc.

Secondly, the current experiments in the realm of WEP are almost always performed resorting to classical systems [10], and the present approach requires the test of WEP employing a quantum particle.

Summing up, the connection with the experiment requires a more careful analysis, and the feasibility of the possible experimental proofs needs a deeper study.  The possible violation of WEP, as a consequence of the presence of a GUP, could seem a very strong drawback of the model. Nevertheless, quantum gravity effects imply the violation of some very well accepted symmetries , for instance, the possible emergence of a deformed dispersion relation in the study of photon propagation [11] (as a consequence, for example, of the polymer--like structure of spacetime) renders the breakdown of Lorentz symmetry [12]. This last remark allows us to contemplate the violation of WEP, stemming from quantum gravity effects, without  any surprise.

Finally, at this point it is noteworthy to comment that the concept of metric theory of gravity (see page 22 of [8]) relies upon the validity of the Weak Equivalence Principle (WEP), and therefore the possible consequences of the present analysis in the context of metric theories could be also interesting . 

\bigskip
\bigskip

\Large{\bf Acknowledgments.}\normalsize
\bigskip

The author would like to thank A. A. Cuevas--Sosa for his help. 

\end{document}